\documentstyle[12pt]{article}
\begin{document}
\parskip=5pt plus 1pt minus 1pt

\begin{flushright}
{\bf UQAM-PHE/95-06}\\
{\bf LMU-12/95}
\end{flushright}

\vspace{0.2cm}
\begin{center}
{\Large\bf A Simplified Approach to Determine $\alpha$ \\
and the Penguin Amplitude in $B_{d}\rightarrow \pi\pi$}
\end{center}

\vspace{0.4cm}
\begin{center}
{\bf Ch$\bf\acute{e}$rif Hamzaoui} \footnote{ Electronic address:
Hamzaoui@mercure.phy.uqam.ca} \\
{\sl D$\sl\acute{e}$partment de Physique, Universit$\sl\acute{e}$ du
Qu$\sl\acute{e}$bec
$\sl\grave{a}$ Montr$\sl\acute{e}$al, \\
Case Postale 8888, Succ. Centre-Ville, Montr$\sl\acute{e}$al,
Qu$\sl\acute{e}$bec,
Canada, H3C 3P8}
\end{center}

\begin{center}
{\bf Zhi-zhong Xing} \footnote{ Electronic address:
Xing@hep.physik.uni-muenchen.de} \\
{\sl Sektion Physik, Theoretische Physik, Universit$\ddot{a}$t
M$\ddot{u}$nchen, \\
Theresienstrasse 37, D-80333 M$\ddot{u}$nchen, Germany}
\end{center}

\vspace{1.2cm}
\begin{abstract}
The effect of inelastic final-state interactions (IFSI's) on the determination
of the weak
phase $\alpha$ from the isospin triangles of $B\rightarrow \pi\pi$ is
qualitatively
illustrated. Neglecting the electroweak penguins and IFSI's and assuming the
dominance of the
top-quark loop in strong penguin diagrams, we propose an experimentally
accessible way to approximately determine $\alpha$ and the penguin amplitude
in $B_{d}\rightarrow \pi\pi$. This approach relies on a simplified isospin
consideration
and the factorization approximation, and its feasibility is irrelevant to the
time-dependent
measurements of $B_d\rightarrow \pi\pi$.
\end{abstract}

\newpage

\begin{flushleft}
{\large\bf 1. Introduction}
\end{flushleft}

Today the most promising way to test the Cabibbo-Kobayashi-Maskawa (CKM)
mechanism of quark
mixing and $CP$ violation is to measure $CP$ asymmetries in neutral $B$-meson
decays
to $CP$ eigenstates \cite{BS}. For this purpose, the decay mode
$B_{d}\rightarrow \pi^{+}\pi^{-}$
is a good candidate in addition to the gold-plated channel $B_{d}\rightarrow
J/\psi K_{S}$.
However, the penguin-induced effect on $B_{d}\rightarrow \pi^{+}\pi^{-}$ may be
significant enough
so that a reliable prediction for the $CP$ asymmetry is theoretically difficult
\cite{PEN}.
Using the isospin triangle relations among the decay amplitudes of
$B^{\pm}_{u}\rightarrow
\pi^{\pm}\pi^{0}$, $\stackrel{(-)}{B}$$^{0}_{d}\rightarrow \pi^{+}\pi^{-}$ and
$\stackrel{(-)}
{B}$$^{0}_{d} \rightarrow \pi^{0}\pi^{0}$, Gronau and London have shown that
the tree-level and
penguin contributions can be disentangled by measuring the relevant decay rates
\cite{GL}.
Their work provides a relatively clean way to extract $\alpha$, an angle of the
CKM
unitarity triangle in the complex plane \cite{CH}, although its feasibility
depends closely upon
the time-dependent measurements of $B_{d}\rightarrow \pi\pi$.

\vspace{0.3cm}

The naive isospin analysis of $B\rightarrow \pi\pi$ has to be modified, if
these decay modes
involve non-negligible electroweak penguin (EWP) diagrams \cite{DH1} or
inelastic final-state
interactions (IFSI's) \cite{FSI}. It has recently been shown that the overall
effect of EWP's on
the determination of $\alpha$ from the isospin triangles is merely at the level
of $O(\lambda^{2})$
($\lambda=0.22$) or smaller \cite{EWP}, although the individual EWP amplitude
may be $O(\lambda)$ of the strong penguin amplitude in $B_d\rightarrow
\pi^0\pi^0$.
The IFSI's are possible to induce channel mixing (e.g., between the
direct reactions $B_d\rightarrow \pi\pi$ and the two-step processes
$B_d\rightarrow D\bar{D}
\rightarrow \pi\pi$), such that both the branching ratios and $CP$ asymmetries
of
$B_d\rightarrow \pi\pi$ are difficult to be predicted. An illustration of the
effect of
IFSI's on the extraction of $\alpha$ from $B\rightarrow \pi\pi$ is still
lacking
in the literacture.

\vspace{0.3cm}

In this work, we first illustrate the influence of $\pi\pi\rightleftharpoons
D\bar{D}$ scattering
on the isospin triangles of $B\rightarrow \pi\pi$ qualitatively. We find that
the
determination of $\alpha$ may suffer from large mixing between the $I=0$ states
of
$\pi\pi$ and $D\bar{D}$. The approach proposed in ref. \cite{GL} can work only
if the IFSI's are insignificant enough to be negligible. Second, we develop an
approximate but more
practical approach to isolate the weak phase $\alpha$ and the penguin amplitude
in $B_{d}\rightarrow \pi\pi$.
Neglecting the effects of EWP's and IFSI's and assuming the dominance of the
top-quark loop
in strong penguin diagrams, we relate the tree-level and penguin amplitudes of
$B\rightarrow \pi\pi$ under a simplified isospin consideration and the
factorization approximation.
Provided that the decay rates of $B^{\pm}_{u}\rightarrow \pi^{\pm}\pi^{0}$,
$\stackrel{(-)}{B}$$^{0}_{d}\rightarrow \pi^{+}\pi^{-}$ and
$\stackrel{(-)}{B}$$^{0}_{d}
\rightarrow \pi^{0}\pi^{0}$ are measured, we show that it is possible to
determine the magnitudes of
the penguin and tree-level amplitudes and to extract the weak and strong phase
shifts between
them. This approximate method has two obvious advantages: (1) its feasibility
depends only upon the
time-independent measurements of the relevant decay rates, which can be carried
out at either
$B$-meson factories or high-luminosity hadron machines; and (2) it can confront
the nearest data of
$B\rightarrow \pi\pi$ and give a ballpark number to be expected for $\alpha$,
before a
delicate determination of $\alpha$ is available in experiments.

\begin{flushleft}
{\large\bf 2. Effect of IFSI's on the isospin triangles}
\end{flushleft}

In the absence of EWP's and IFSI's, the isospin amplitudes of $B^0_d\rightarrow
\pi^+\pi^-$,
$B^+_u\rightarrow \pi^+\pi^0$, $B^0_d\rightarrow \pi^0\pi^0$ and their
$CP$-conjugate
counterparts are given as \cite{GL}
$$
\begin{array}{ccccl}
A_{+-} & = & \langle \pi^{+}\pi^{-}|{\cal H}|B^{0}_{d}\rangle & = &
\sqrt{2} A^{\pi\pi}_2 ~ - ~ \sqrt{2} A^{\pi\pi}_0 \; , \\
A_{+0} & = & \langle \pi^{+}\pi^{0}|{\cal H}|B^{+}_{u}\rangle & = & 3
A^{\pi\pi}_2 \; , \\
A_{00} & = & \langle \pi^{0}\pi^{0}|{\cal H}|B^{0}_{d}\rangle & = & 2
A^{\pi\pi}_2 ~ + ~ A^{\pi\pi}_0 \; ;
\end{array}
\eqno(1{\rm a})
$$
and
$$
\begin{array}{ccccl}
\bar{A}_{+-} & = & \langle \pi^{+}\pi^{-}|{\cal H}|\bar{B}^{0}_{d}\rangle & = &
\sqrt{2} \bar{A}^{\pi\pi}_2 ~ - ~ \sqrt{2} \bar{A}^{\pi\pi}_0 \; , \\
\bar{A}_{-0} & = & \langle \pi^{-}\pi^{0}|{\cal H}|B^{-}_{u}\rangle & = & 3
\bar{A}^{\pi\pi}_2 \; , \\
\bar{A}_{00} & = & \langle \pi^{0}\pi^{0}|{\cal H}|\bar{B}^{0}_{d}\rangle & = &
2 \bar{A}^{\pi\pi}_2 ~ + ~ \bar{A}^{\pi\pi}_0 \; ,
\end{array}
\eqno(1{\rm b})
$$
where some Clebsch-Gordan coefficients have been absorbed into the definitions
of
$A^{\pi\pi}_{0,2}$ and $\bar{A}^{\pi\pi}_{0,2}$. Clearly the above relations
form two
isospin triangles in the complex plane:
$$
A_{+-} + \sqrt{2} A_{00} \; =\; \sqrt{2} A_{+0} \; , ~~~~~~~~~~
\bar{A}_{+-} + \sqrt{2} \bar{A}_{00} \; =\; \sqrt{2} \bar{A}_{-0} \; .
\eqno(2)
$$
Since the decay modes $B^{\pm}_u\rightarrow \pi^{\pm}\pi^0$ occur solely
through the tree-level
quark diagrams, the ratio $\bar{A}^{\pi\pi}_2 / A^{\pi\pi}_2 = \bar{A}_{-0} /
A_{+0}$
is purely governed by the CKM factor ($V_{ub}V^*_{ud})/(V^*_{ub}V_{ud}) =
e^{-2i\gamma}$
($\gamma$ is an angle of the unitarity triangle \cite{PDG}). To probe $CP$
violation
induced by the interplay of decay and $B^0_d -\bar{B}^0_d$ mixing in
$B_d\rightarrow
\pi^+\pi^-$, we need to measure the interference term
$$
{\rm Im}\xi_{+-} \; =\; {\rm Im} \left ( e^{-2i\beta} ~
\frac{\bar{A}_{+-}}{A_{+-}}\right ) \; =\;
{\rm Im} \left (e^{2i\alpha} ~ \frac{1-\bar{z}}{1-z}\right ) \; ,
\eqno(3)
$$
where $\beta$ and $\alpha$ are also the angles of the CKM unitarity triangle
($\alpha + \beta +\gamma =180^0$), $z=A^{\pi\pi}_0 / A^{\pi\pi}_2$ and
$\bar{z}=\bar{A}^{\pi\pi}_0 / \bar{A}^{\pi\pi}_2$. As shown in ref. \cite{GL},
the complex parameter $z$ ($\bar{z}$) is determinable from the isospin
triangles up to a two-fold ambiguity
in the sign of its phase. In a similar way one can discuss the $CP$-violating
term
${\rm Im}\xi_{00}$ in $B_d\rightarrow \pi^0\pi^0$. Then the weak angle $\alpha$
can be extracted
from ${\rm Im}\xi_{+-}$ and ${\rm Im}\xi_{00}$. This approach relies on the
time-dependent
measurements of $B_d\rightarrow \pi\pi$.

\vspace{0.3cm}

Now let us illustrate the effect of IFSI's on the isospin triangles and the
$CP$-violating
measurable ${\rm Im}\xi_{+-}$ (${\rm Im}\xi_{00}$), which are crucial for the
extraction of $\alpha$.
Due to IFSI's the $I=0$ state of $\pi\pi$ can mix with that of $D\bar{D}$, and
this leads to the
two-step decays $B_d\rightarrow D\bar{D} \rightarrow \pi\pi$. The final states
$\pi^{\pm}\pi^0$
cannot mix with $D\bar{D}$ because of their different isospin configurations
($I=2$ for $\pi^{\pm}\pi^0$, $I=0$ and $I=1$ for $D\bar{D}$). But in principle
$\pi^{\pm}\pi^0$
could mix with $\rho^{\pm}\rho^0$, and $D\bar{D}$ could mix with $D^*\bar{D}^*$
\cite{FSI}.
For simplicity, here we only consider the influence of
$\pi\pi\rightleftharpoons D\bar{D}$
scattering on $B_d\rightarrow \pi\pi$, leaving the $I=2$ state of $\pi\pi$
unmixed with
others. As a result, the ``bare'' isospin amplitudes of $B_d\rightarrow \pi\pi$
in eq. (1) are
modified as
$$
\begin{array}{ccl}
A^{'}_{+-} & = & \sqrt{2} S^{\pi\pi}_2 A^{\pi\pi}_2 ~ - ~ \sqrt{2} \left (
S^{\pi\pi}_0 A^{\pi\pi}_0
+ S^{\pi D}_0 A^{DD}_0 \right ) \; , \\
A^{'}_{00} & = & 2 S^{\pi\pi}_2 A^{\pi\pi}_2 ~ + ~ \left ( S^{\pi\pi}_0
A^{\pi\pi}_0 + S^{\pi D}_0 A^{DD}_0
\right ) \; ;
\end{array}
\eqno(4{\rm a})
$$
and
$$
\begin{array}{ccl}
\bar{A}^{'}_{+-} & = & \sqrt{2} S^{\pi\pi}_2 \bar{A}^{\pi\pi}_2 ~ - ~ \sqrt{2}
\left (
S^{\pi\pi}_0 \bar{A}^{\pi\pi}_0 + S^{\pi D}_0 \bar{A}^{DD}_0 \right ) \; , \\
\bar{A}^{'}_{00} & = & 2 S^{\pi\pi}_2 \bar{A}^{\pi\pi}_2 ~ + ~ \left (
S^{\pi\pi}_0 \bar{A}^{\pi\pi}_0
+ S^{\pi D}_0 \bar{A}^{DD}_0 \right ) \; .
\end{array}
\eqno(4{\rm b})
$$
In the above equations, $A^{DD}_0$ denote the $I=0$ amplitude component of
$B_d\rightarrow D\bar{D}$,
$S_{0,2}$ are complex matrix elements connecting the unitarized isospin
amplitudes to the bare ones
\cite{FSI}. Of course, $S^{\pi D}_0 =0$ and $S^{\pi\pi}_2 = S^{\pi\pi}_0 =1$ if
there is no
mixture between the $I=0$ states of $\pi\pi$ and $D\bar{D}$. From eq. (4), we
obtain the
following triangle relations:
$$
A^{'}_{+-} + \sqrt{2} A^{'}_{00} \; =\; \sqrt{2} S^{\pi\pi}_2 A_{+0} \; ,
{}~~~~~~~~~~
\bar{A}^{'}_{+-} + \sqrt{2} \bar{A}^{'}_{00} \; =\; \sqrt{2} S^{\pi\pi}_2
\bar{A}_{-0} \; .
\eqno(5)
$$
Comparing this result with eq. (2), one can observe an overall change of the
bare isospin triangles due to IFSI's.
In this case, the $CP$-violating measurable ${\rm Im}\xi_{+-}$ becomes
$$
{\rm Im}\xi^{'}_{+-} \; =\; {\rm Im}\left (e^{-2i\beta} ~
\frac{\bar{A}^{'}_{+-}}{A^{'}_{+-}}\right ) \; =\;
{\rm Im} \left (e^{2i\alpha} ~ \frac{1-\bar{z}'}{1-z'}\right ) \; ,
\eqno(6)
$$
where
$$
z' \; =\; z ~ \frac{S^{\pi\pi}_0}{S^{\pi\pi}_2} ~ + ~
\frac{A^{DD}_0}{A^{\pi\pi}_2}
{}~ \frac{S^{\pi D}_0}{S^{\pi\pi}_2} \; , ~~~~~~~~~
\bar{z}' \; =\; \bar{z} ~ \frac{S^{\pi\pi}_0}{S^{\pi\pi}_2} ~ + ~
\frac{\bar{A}^{DD}_0}
{\bar{A}^{\pi\pi}_2} ~ \frac{S^{\pi D}_0}{S^{\pi\pi}_2} \; .
\eqno(7)
$$
We find that in principle the angle $\alpha$ can still be extracted from
${\rm Im}\xi^{'}_{+-}$ (and ${\rm Im}\xi^{'}_{00}$). However, the determination
of $z'$ and $\bar{z}'$
from eq. (5) needs the knowledge of $S^{\pi\pi}_2$, since the measured
branching ratio of
$B^{+}_u\rightarrow \pi^{+}\pi^0$ can only fix $|A_{+0}|$ other than
$|S^{\pi\pi}_2 A_{+0}|$. Before we obtain definite information on the
mixing of $\pi\pi$ and $D\bar{D}$, it is impossible to determine the values of
$|S^{\pi D}_0|$ and $|S^{\pi\pi}_{0,2}|$ reliably
\footnote{From ref. \cite{FSI}, one can obtain a rough feeling of the possible
effect of
IFSI's on the magnitudes of decay rates and $CP$ asymmetries.}. Thus the
isospin approach of ref. \cite{GL}
cannot work well in the presence of significant IFSI's.

\vspace{0.3cm}

If the $\pi\pi\rightleftharpoons D\bar{D}$ scattering effect is insignificant
in the
problem under discussion, one expects that $|S^{\pi D}_0|$ may be small enough
and
$|S^{\pi\pi}_2|$ does not deviate too much from unity. In this case,
the two triangles in eq. (5) are able to be approximately determined from
direct measurements,
then one can isolate the unknow parameters $z'$ and $\bar{z}'$ to an acceptable
degree of accuracy and extract $\alpha$ from ${\rm Im}\xi^{'}_{+-}$ and ${\rm
Im}\xi^{'}_{00}$.
In the following we shall assume this simple case and present an approximate
approach to determine $\alpha$ and the penguin amplitude in the decay modes
$B_d\rightarrow \pi\pi$.

\begin{flushleft}
{\large\bf 3. Approximate isolation of the penguin amplitude}
\end{flushleft}

To lowest order in weak interactions, all two-body mesonic $B$ decays can be
generally described
by ten topologically different quark diagrams \cite{XDiag}. In this language,
we assume that
the effect of IFSI's on every decay mode is insignificant enough to be
negligible.
Neglecting the EWP diagrams and those exchange- or annihilation-type channels,
the dominant quark-diagram
amplitudes for $B^{0}_{d}\rightarrow \pi^{+}\pi^{-}$, $B^{+}_{u}\rightarrow
\pi^{+}\pi^{0}$ and
$B^{0}_{d}\rightarrow \pi^{0}\pi^{0}$ are illustrated in fig. 1. Another
reasonable assumption to be used
is that the strong penguin diagram is dominated by the top-quark loop
\cite{W,GRL}. This implies that the
weak phase of the penguin amplitude comes only from $V^{*}_{tb}V_{td}$ in
$B_{d}^{0}\rightarrow
\pi^{+}\pi^{-}$ or $B^{0}_{d}\rightarrow \pi^{0}\pi^{0}$. In terms of the
angles of the
CKM unitarity triangle \cite{PDG}, we parametrize the decay amplitudes of the
three modes
and their $CP$-conjugate counterparts as follows
\footnote{Here we use the same valence-quark notations for the pions as refs.
\cite{Z,GHLR}:
$|\pi^+\rangle = |u\bar{d}\rangle$, $|\pi^0\rangle = |d\bar{d} -u\bar{u}\rangle
/\sqrt{2}$,
and $|\pi^-\rangle = -|d\bar{u}\rangle$. This convention  guarantees that the
three pion mesons belong
to an isotriplet. Also, we adopt $|B^0_d\rangle =|d\bar{b}\rangle$,
$|\bar{B}^0_d\rangle =
|b\bar{d}\rangle$, $|B^+_u\rangle =|u\bar{b}\rangle$, and $|B^-_u\rangle =
-|b\bar{u}\rangle$.}:
$$
\begin{array}{ccl}
A_{+-} & = &  - T e^{i\gamma} - P e^{i(\delta -\beta)} \; , \\
A_{+0} & = &  - \displaystyle\frac{1+a}{\sqrt{2}} T e^{i\gamma} \; , \\
A_{00} & = &  - \displaystyle\frac{a}{\sqrt{2}}T e^{i\gamma}
+ \displaystyle\frac{1}{\sqrt{2}}P e^{i(\delta - \beta)} \; ;
\end{array}
\eqno(8{\rm a})
$$
and
$$
\begin{array}{ccl}
\bar{A}_{+-} & = & - T e^{-i\gamma} - P e^{i(\delta +\beta)} \; , \\
\bar{A}_{-0} & = & - \displaystyle\frac{1+a}{\sqrt{2}} T e^{-i\gamma} \; , \\
\bar{A}_{00} & = & - \displaystyle\frac{a}{\sqrt{2}}
T e^{-i\gamma} + \displaystyle\frac{1}{\sqrt{2}}P e^{i(\delta + \beta)} \; .
\end{array}
\eqno(8{\rm b})
$$
The above expressions are based on a simple isospin consideration and the
factorization approximation
for the final states $\pi\pi$, as we shall explain below. Here $\delta$ denotes
the strong phase
shift between the penguin and tree-level amplitudes
of $B_{d}\rightarrow \pi\pi$. Thus without loss of any generality, $T$ and $P$
can be assumed to be
real and positive. The parameter $a$ represents the color-mismatched
suppression in the tree-level
amplitudes of $B^{+}_{u}\rightarrow \pi^{+}\pi^{0}$, $B^{0}_{d}\rightarrow
\pi^{0}\pi^{0}$ and their
$CP$-conjugate processes. We shall see later that $a$ can be either calculated
using the factorization
approximation or measured directly from the decay rates. Clearly the relations
in eq. (8)
can form the same isospin triangles as those in eq. (2).
This implies that the quark-diagram language used for describing $B\rightarrow
\pi\pi$ is
consistent with the isospin analysis, as shown first in ref. \cite{GHLR}.
In the factorization approximation, we can assume that the difference between
the interactions of
$I=0$ and $I=2$ final states is negligible. Thus the two quark-diagram
amplitudes of
$B^{+}_{u}\rightarrow \pi^{+}\pi^{0}$, which occurs solely through the $\Delta
I=3/2$ transition,
can be parametrized as eq. (8a) with a real factor $a$. This accordingly
implies that the tree-level amplitude
of $B^{0}_{d}\rightarrow \pi^{+}\pi^{-}$ and that of $B^{0}_{d}\rightarrow
\pi^{0}\pi^{0}$
also have negligible strong phase shift, to guarantee the isospin triangle
relations in eq. (2).
So do the penguin amplitudes of $B^{0}_{d}\rightarrow \pi^{+}\pi^{-}$ and
$B^{0}_{d}\rightarrow
\pi^{0}\pi^{0}$. Therefore, we expect that contributions from the penguin
diagrams of
$B_{d}\rightarrow \pi\pi$ are crucial for the isospin triangles to have
nonvanishing area.
For a detailed discussion about translation of the isospin amplitudes of
$B\rightarrow
\pi\pi$ into the corresponding combinations of quark-diagram amplitudes, we
refer the reader to
ref. \cite{GHLR}. Based on the correspondence between the quark-diagram
description and the isospin analysis,
we subsequently use eq. (8) to relate the weak and strong phases in
$B_{d}\rightarrow \pi\pi$ to the
time-independent measurables.

\vspace{0.3cm}

{}From eqs. (8a) and (8b), we have $|A_{+0}|=|\bar{A}_{-0}|$.
The decay rates of the above six processes should be measured in the near
future at either $B$-meson
factories or high-luminosity hadron machines. Let us define the following four
measurables:
$$
R_{+-} \; =\; \frac{|A_{+-}|^{2}}{|A_{+0}|^{2} +|\bar{A}_{-0}|^{2}} \; ,
{}~~~~~~~~
\bar{R}_{+-}\; =\; \frac{|\bar{A}_{+-}|^{2}}{|A_{+0}|^{2} +|\bar{A}_{-0}|^{2}}
\; ;
\eqno(9{\rm a})
$$
$$
R_{00} \; =\; \frac{|A_{00}|^{2}}{|A_{+0}|^{2} +|\bar{A}_{-0}|^{2}} \; ,
{}~~~~~~~~
\bar{R}_{00}\; =\; \frac{|\bar{A}_{00}|^{2}}{|A_{+0}|^{2} +|\bar{A}_{-0}|^{2}}
\; .
\eqno(9{\rm b})
$$
It is clear that $T$ can be directly obtained from $|A_{+0}|^{2}$ or
$|\bar{A}_{-0}|^{2}$.
On the other hand, the magnitude of the penguin amplitude $P$ is only
determined once the data
on $R_{+-}$ ($\bar{R}_{+-}$) and $R_{00}$ ($\bar{R}_{00}$) become available.
By use of eqs. (8) and (9), the ratio $P/T\equiv \chi$ is given as
$$
\chi \; = \; \sqrt{ (1+a) \left (a R_{+-} + 2 R_{00}\right ) -a } \; =\;
\sqrt{ (1+a) \left (a \bar{R}_{+-} + 2 \bar{R}_{00}\right ) -a } \; .
\eqno(10)
$$
It should be noted that the size of $\chi$ is sensitive to the uncertainty
associated with the
color-suppression factor $a$. There are two ways to determine $a$. One way is
to use the
effective weak Hamiltonian ${\cal H}(\Delta B=1)$ \cite{Bu1} and the
factorization
approximation \cite{Bu2}. It is easy to obtain
$$
a\; =\; \frac{C_{2}+N_{c}C_{1}}{C_{1}+N_{c}C_{2}} \; ,
\eqno(11)
$$
where $N_{c}$ is the number of colors, $C_{1}\approx -0.291$ and $C_{2}\approx
1.133$ are the
Wilson coefficients at the scale $\mu=O(m_{b})$ \cite{Bu1}. We obtain $a\approx
0.23$ for
$N_{c}\approx 2.2$ and $a\approx -0.26$ for $N_{c}=\infty$. The current data on
exclusive
hadronic $B$ decays favor the former, i.e. a positive $a$ with the magnitude of
$O(\lambda)$
\cite{CLEO}. The other way to determine $a$ is
through measurements of the decay modes $B_{d}\rightarrow \pi^{+}\pi^{-}$ and
$B_{d}\rightarrow \pi^{0}\pi^{0}$. From eqs. (8) and (9) one can find
$$
a \; =\; -2\cdot \frac{\bar{R}_{00}-R_{00}}{\bar{R}_{+-}-R_{+-}} \; .
\eqno(12)
$$
Indeed the rate differences $\bar{R}_{00}-R_{00}$ and $\bar{R}_{+-}-R_{+-}$
indicate the
signals of direct $CP$ violation. The assumptions made in eq. (8) imply that
the direct $CP$
asymmetries in $B_{d}\rightarrow \pi^{+}\pi^{-}$ and $B_{d}\rightarrow
\pi^{0}\pi^{0}$ are
governed by the same weak phase difference ($\beta +\gamma$) and the same
strong phase shift
($\delta$), but their magnitudes are different.
A comparison between the results of $a$ obtained in eqs. (11) and (12) can
serve as a rough
test of the factorization approximation. In practice, however, measuring the
rate difference
$\bar{R}_{00}-R_{00}$ might be very difficult due to the expected smallness of
the branching ratio of
$B_{d}\rightarrow \pi^{0}\pi^{0}$. Considering $SU(3)$ symmetry and its small
breaking effect,
Deshpande and He \cite{DH} have shown that the rate difference of
$B^{0}_{d}\rightarrow
\pi^{+}\pi^{-}$ ($\pi^{0}\pi^{0}$) vs $\bar{B}^{0}_{d}\rightarrow
\pi^{+}\pi^{-}$ ($\pi^{0}\pi^{0}$) is related to that of $B^{0}_{d}\rightarrow
\pi^{-}K^{+}$
($\pi^{0}K^{0}$) vs $\bar{B}^{0}_{d} \rightarrow \pi^{+}K^{-}$
($\pi^{0}\bar{K}^{0}$). The
latter can be more easily measured in experiments, thus can be used to
determine $a$ through
eq. (12). It should be noted that our main results still hold even in the
special case $a=0$.
If $\delta $ happens to be $0^{0}$ or $180^{0}$, however, eq. (12) will not
work. In this case,
we have to use the data on other $B$ decays to determine $a$ \cite{CLEO}.

\begin{flushleft}
{\large\bf 4. Approximate determination of $\alpha$}
\end{flushleft}

Now we look at how to extract the weak phase $\alpha$ without resorting to
measurements of the
decay-time distribution of $B_{d}\rightarrow \pi\pi$. Since the weak phase
shift between the
penguin and tree-level amplitudes of $B_{d}\rightarrow \pi^{+}\pi^{-}$ is
$\beta +\gamma$,
by use of the unitarity condition $\alpha +\beta +\gamma =180^{0}$ we obtain
two concise equations
connecting the weak and strong phases to the measurables:
$$
\cos (\alpha +\delta ) \; = \; \frac{1}{2\chi}\left [ 1+\chi^{2} - (1+a)^{2}
R_{+-}\right ] \; ,
\eqno(13{\rm a})
$$
$$
\cos (\alpha -\delta ) \; = \; \frac{1}{2\chi}\left [ 1+\chi^{2} - (1+a)^{2}
\bar{R}_{+-}\right ] \; .
\eqno(13{\rm b})
$$
Similarly, $\alpha$ and $\delta$ can be related to $R_{00}$ and $\bar{R}_{00}$.
Because
$\chi$ is determined from eq. (10), both $\alpha$ and $\delta$ can be extracted
from (13)
with four-fold ambiguity. Due to the assumptions made in eq. (8), this
ambiguity cannot be
reduced by the relations among $\alpha$, $\delta$, $R_{00}$ and $\bar{R}_{00}$.
To the
accuracy that $R_{+-}$ and $\bar{R}_{+-}$ cannot be distinguished in
experiments, we
expect that the strong phase shift $\delta$ might be close to $0^{0}$ or
$180^{0}$ (i.e.,
vanishingly small direct $CP$ violation in $B_{d}\rightarrow \pi\pi$). Note
that eq. (13)
would become invalid if the magnitude of the penguin amplitude $P$ were too
small to be
measured. In this case, the two isospin triangles in eq. (2) would collapse to
lines.
Within the standard model, however, several calculations have given that $\chi$
is at the
level of $O(\lambda)$ or larger \cite{PEN}. Thus our results (10) and (13) are
feasible to
probe the weak phase $\alpha$ as well as the significant penguin amplitude in
$B_{d}\rightarrow \pi\pi$.

\vspace{0.3cm}

It is worth remarking that our approach to determine the angle $\alpha$ does
not require the measurement of
the $CP$-violating term ${\rm Im}\xi_{+-}$:
$$
{\rm Im}\xi_{+-} \; \approx \; \sin (2\alpha) - 2\chi \sin\alpha \cos\delta
\eqno(14)
$$
(to lowest-order approximation of $\chi$), which is induced by the interplay of
decay
and $B^{0}_{d}-\bar{B}^{0}_{d}$ mixing. Thus the feasibility of our approach is
independent of
the time-dependent measurements of neutral $B$ decays to be carried out at the
asymmetric
$B$ factories of KEK and SLAC \cite{BFAC}.
Nonetheless, it should be noted that the determination of $|A_{+-}|$ and
$|\bar{A}_{+-}|$ from any
experiment is indeed influenced by $B^{0}_{d}-\bar{B}^{0}_{d}$ mixing. To
overcome this problem,
one available way is through the time-integrated measurements of
$B_{d}\rightarrow
\pi^{+}\pi^{-}$ on the $\Upsilon (4S)$ resonance, where the produced two
$B_{d}$ mesons are in a
coherent state (with odd charge-conjugation parity) until one of them decays.
Using the semileptonic
decay of one $B_{d}$ meson to tag the flavor of the other meson decaying to
$\pi^{+}\pi^{-}$, the
probability for observing such a joint decay has been give as \cite{4S}:
$$
{\rm Pr}(l^{+}X^{-}; \pi^{+}\pi^{-}) \; =\; |A_{l}|^{2} \left [
\frac{|A_{+-}|^{2} + |\bar{A}_{+-}|^{2}}{2} - \frac{1}{1+x^{2}_{d}}\cdot
\frac{|A_{+-}|^{2}
-|\bar{A}_{+-}|^{2}}{2} \right ] \; ,
\eqno(15{\rm a})
$$
$$
{\rm Pr}(l^{-}X^{+}; \pi^{+}\pi^{-}) \; =\; |A_{l}|^{2} \left [
\frac{|A_{+-}|^{2} + |\bar{A}_{+-}|^{2}}{2} + \frac{1}{1+x^{2}_{d}}\cdot
\frac{|A_{+-}|^{2}
-|\bar{A}_{+-}|^{2}}{2} \right ] \; ,
\eqno(15{\rm b})
$$
where $|A_{l}|=|\langle l^{+}X^{-}|{\cal H}|B^{0}_{d}\rangle |
\stackrel{CPT}{=}
|\langle l^{-}X^{+}|{\cal H}|\bar{B}^{0}_{d}\rangle |$, and $x_{d}=\Delta
m/\Gamma\approx 0.71$ is
a measure of $B^{0}_{d}-\bar{B}^{0}_{d}$ mixing \cite{PDG}.
At present, the semileptonic decays $B^{0}_{d}\rightarrow l^{+}X^{-}$ and
$\bar{B}^{0}_{d}
\rightarrow l^{-}X^{+}$ have been well reconstructed \cite{PDG,ST}, thus
$|A_{l}|$ is
determinable independent of the above joint decays. Once ${\rm
Pr}(l^{\pm}X^{\mp}; \pi^{+}\pi^{-})$
are measured, we shall be able to determine $|A_{+-}|$ and $|\bar{A}_{+-}|$
through eq. (15).
In practice, the measurements of ${\rm Pr}(l^{\pm}X^{\mp}; \pi^{+}\pi^{-})$ can
be realized
either by the symmetric $e^{+}e^{-}$ collider running at Cornell \cite{ST} or
by the forthcoming
asymmetric $B$ factories. Similarly, $|A_{00}|$ and $|\bar{A}_{00}|$ can be
determined.

\begin{flushleft}
{\large\bf 5. Discussions and conclusion}
\end{flushleft}

In the first part of this work, we have qualitatively illustrated the effect of
IFSI's on
the isospin triangles of $B\rightarrow \pi\pi$. The determination of $\alpha$
from these
decay modes may suffer from large mixing between the $I=0$ states of $\pi\pi$
and
$D\bar{D}$. A quantitative estimation of IFSI's needs much progress in theory
and more data in experiments.

\vspace{0.3cm}

In the assumption that the IFSI's in $B\rightarrow \pi\pi$ are insignificant
enough to be negligible, we
have presented a self-consistent and time-independent approach to approximately
isolate
the weak phase $\alpha$ and the penguin amplitude of $B_{d}\rightarrow \pi\pi$.
Here
it is worthwhile to comment briefly on the uncertainties with this approach. In
the first approximation, we neglected
the electroweak penguins in $B\rightarrow \pi\pi$. As pointed out in ref.
\cite{EWP},
the effects of electroweak penguins on the decays via $\bar{b}\rightarrow
\bar{u}u\bar{d}$ or
$\bar{b}\rightarrow \bar{d}$ are at most $O(\lambda^{2})$ of the dominant
tree-level amplitude $T$
or $O(\lambda)$ of the strong penguin amplitude $P$.
Hence it is unlikely that the validity of the isospin analysis in ref.
\cite{GL} and our present method
are significantly affected by the neglect of the electroweak penguins.
On the other hand, the approximation of neglecting those exchange- and
annihilation-type
topologies in $B\rightarrow \pi\pi$ is expected to be safe for some dynamical
reasons \cite{GHLR}.
The approach to test the reliability of this kind of approximations has been
suggested in ref. \cite{GRL}.
The factorization approximation used to parametrize the decay amplitudes of
$B\rightarrow \pi\pi$
as eq. (8) is equivalent to neglecting the difference between the $I=0$ and
$I=2$ interactions in the final
states. The validity of this assumption can be examined by the forthcoming
experimental data. The assumption that the strong
penguin diagram is dominated by the top-quark loop can be interpreted as
follows. On the basis of the
standard model with three families of quarks, the penguin contributions with
$u$ and $c$ quarks
in the loop are equal in magnitude up to $O(m^{2}_{c}/m^{2}_{b})$ apart from
their CKM factors.
Thus the top-quark loop dominates the penguin amplitude to the same degree of
accuracy, i.e.
$O(\leq 10\%)$, guaranteed by the CKM unitarity. Indeed this is a good
approximation to
obtain the lowest order magnitudes of $\alpha$, $\delta$ and $P$. When the
time-dependent measurements
of $B_{d}\rightarrow \pi\pi$ become available at an asymmetric $B$ factory to
allow for the more
precise isospin analysis proposed by Gronau and London \cite{GL}, one can check
the difference between
our results and theirs. The corresponding deviation should definitely constrain
the reliability of the
top-quark dominance and the factorization approximation assumed in our work.

\vspace{0.3cm}

By now the experimental sensitivity to the penguin contribution in
$B_{d}\rightarrow \pi\pi$ has been
analyzed based on the isospin triangle relations \cite{GL} and under the
circumstance of asymmetric
$B$-meson factories \cite{A}. Such an analysis is also applicable to the
simplified approach
proposed in this work. It is concluded that determining the weak and strong
phases associated with
the penguin amplitude to a satisfactory accuracy is experimentally feasible in
the near future.
Finally we stress again that our approach is time-independent, and thus it can
be confronted with all
the forthcoming measurements of $B\rightarrow \pi\pi$, no matter whether they
are time-integrated or
time-dependent.

\begin{flushleft}
{\large\bf Acknowledgements}
\end{flushleft}

We are very grateful to X.G. He, T. Nakada and S. Stone for useful discussions
and comments.
Z.Z.X. would like to thank H. Fritzsch for his warm hospitality and constant
encouragement. C.H. Would like to thank N. Khuri and the Physics Department of
Rockefeller University, where part of this work was done, for their warm
hospitality.
The work of C.H. was partially funded by the N.S.E.R.C. of Canada, and the
research
of Z.Z.X. was supported by the A.v.H. Foundation of Germany.

\newpage

\small
\begin{figure}
\begin{picture}(400,250)(0,0)
\put(70,215){\line(1,0){90}}
\put(62,213){$\bar{b}$}
\put(62,186){$d$}
\put(43,200){$B^{0}_{d}$}
\put(163,212){$\bar{u}$}
\put(163,186){$d$}
\put(163,244){$\bar{d}$}
\put(163,230){$u$}
\put(174,199){$\pi^{-}$}
\put(174,236){$\pi^{+}$}
\put(70,190){\line(1,0){90}}
\put(160,240){\oval(70,15)[l]}
\put(145,247.5){\vector(-1,0){2}}
\put(145,232.5){\vector(1,0){2}}
\put(85,215){\vector(-1,0){2}}
\put(145,215){\vector(-1,0){2}}
\put(85,190){\vector(1,0){2}}
\put(145,190){\vector(1,0){2}}
\multiput(110,215)(3,5){5}{\line(0,1){5}}
\multiput(107,215)(3,5){6}{\line(1,0){3}}
\put(113,165){(a1)}
\put(270,240){\line(1,0){25}}
\put(325,240){\line(1,0){35}}
\put(310,240){\oval(30,36)[b]}
\multiput(294,240)(5,0){6}{$\wedge$}
\put(270,190){\line(1,0){90}}
\put(262,238){$\bar{b}$}
\put(262,186){$d$}
\put(243,210.5){$B^{0}_{d}$}
\put(363,238){$\bar{d}$}
\put(363,186){$d$}
\put(363,226){$u$}
\put(363,200){$\bar{u}$}
\put(374,193){$\pi^{-}$}
\put(374,230){$\pi^{+}$}
\put(360,215){\oval(60,25)[l]}
\put(285,240){\vector(-1,0){2}}
\put(285,190){\vector(1,0){2}}
\put(345,240){\vector(-1,0){2}}
\put(345,190){\vector(1,0){2}}
\put(345,227.5){\vector(1,0){2}}
\put(345,202.5){\vector(-1,0){2}}
\put(313,165){(a2)}
\end{picture}

\begin{picture}(400,150)(0,0)
\put(70,215){\line(1,0){90}}
\put(62,213){$\bar{b}$}
\put(62,186){$u$}
\put(43,200){$B^{+}_{u}$}
\put(163,212){$\bar{u}$}
\put(163,186){$u$}
\put(163,244){$\bar{d}$}
\put(163,230){$u$}
\put(174,199){$\pi^{0}$}
\put(174,236){$\pi^{+}$}
\put(70,190){\line(1,0){90}}
\put(160,240){\oval(70,15)[l]}
\put(145,247.5){\vector(-1,0){2}}
\put(145,232.5){\vector(1,0){2}}
\put(85,215){\vector(-1,0){2}}
\put(145,215){\vector(-1,0){2}}
\put(85,190){\vector(1,0){2}}
\put(145,190){\vector(1,0){2}}
\multiput(110,215)(3,5){5}{\line(0,1){5}}
\multiput(107,215)(3,5){6}{\line(1,0){3}}
\put(113,165){(b1)}
\put(270,240){\line(1,0){90}}
\put(270,190){\line(1,0){90}}
\put(262,238){$\bar{b}$}
\put(262,186){$u$}
\put(243,210.5){${B}^{+}_{u}$}
\put(363,238){$\bar{u}$}
\put(363,186){$u$}
\put(363,223){$u$}
\put(363,200){$\bar{d}$}
\put(374,193){$\pi^{+}$}
\put(374,230){$\pi^{0}$}
\put(360,215){\oval(70,25)[l]}
\put(285,240){\vector(-1,0){2}}
\put(285,190){\vector(1,0){2}}
\put(345,240){\vector(-1,0){2}}
\put(345,190){\vector(1,0){2}}
\put(345,227.5){\vector(1,0){2}}
\put(345,202.5){\vector(-1,0){2}}
\multiput(310,240)(3,-5){5}{\line(0,-1){5}}
\multiput(307,240)(3,-5){6}{\line(1,0){3}}
\put(313,165){(b2)}
\end{picture}

\begin{picture}(400,150)(0,0)
\put(70,240){\line(1,0){90}}
\put(70,190){\line(1,0){90}}
\put(62,238){$\bar{b}$}
\put(62,186){$d$}
\put(43,210.5){$B^{0}_{d}$}
\put(163,238){$\bar{u}$}
\put(163,186){$d$}
\put(163,223){$u$}
\put(163,200){$\bar{d}$}
\put(174,193){$\pi^{0}$}
\put(174,230){$\pi^{0}$}
\put(160,215){\oval(70,25)[l]}
\put(85,240){\vector(-1,0){2}}
\put(85,190){\vector(1,0){2}}
\put(145,240){\vector(-1,0){2}}
\put(145,190){\vector(1,0){2}}
\put(145,227.5){\vector(1,0){2}}
\put(145,202.5){\vector(-1,0){2}}
\multiput(110,240)(3,-5){5}{\line(0,-1){5}}
\multiput(107,240)(3,-5){6}{\line(1,0){3}}
\put(113,165){(c1)}
\put(270,240){\line(1,0){25}}
\put(325,240){\line(1,0){35}}
\put(310,240){\oval(30,36)[b]}
\multiput(294,240)(5,0){6}{$\wedge$}
\put(270,190){\line(1,0){90}}
\put(262,238){$\bar{b}$}
\put(262,186){$d$}
\put(243,210.5){$B^{0}_{d}$}
\put(363,238){$\bar{d}$}
\put(363,186){$d$}
\put(363,223){$d$}
\put(363,200){$\bar{d}$}
\put(374,193){$\pi^{0}$}
\put(374,230){$\pi^{0}$}
\put(360,215){\oval(60,25)[l]}
\put(285,240){\vector(-1,0){2}}
\put(285,190){\vector(1,0){2}}
\put(345,240){\vector(-1,0){2}}
\put(345,190){\vector(1,0){2}}
\put(345,227.5){\vector(1,0){2}}
\put(345,202.5){\vector(-1,0){2}}
\put(313,165){(c2)}
\end{picture}
\vspace{-4.8cm}
\caption{The dominant tree-level and strong penguin diagrams for
$B^{0}_{d}\rightarrow \pi^{+}\pi^{-}$,
$B^{+}_{u}\rightarrow \pi^{+}\pi^{0}$, and $B^{0}_{d}\rightarrow
\pi^{0}\pi^{0}$.}
\end{figure}
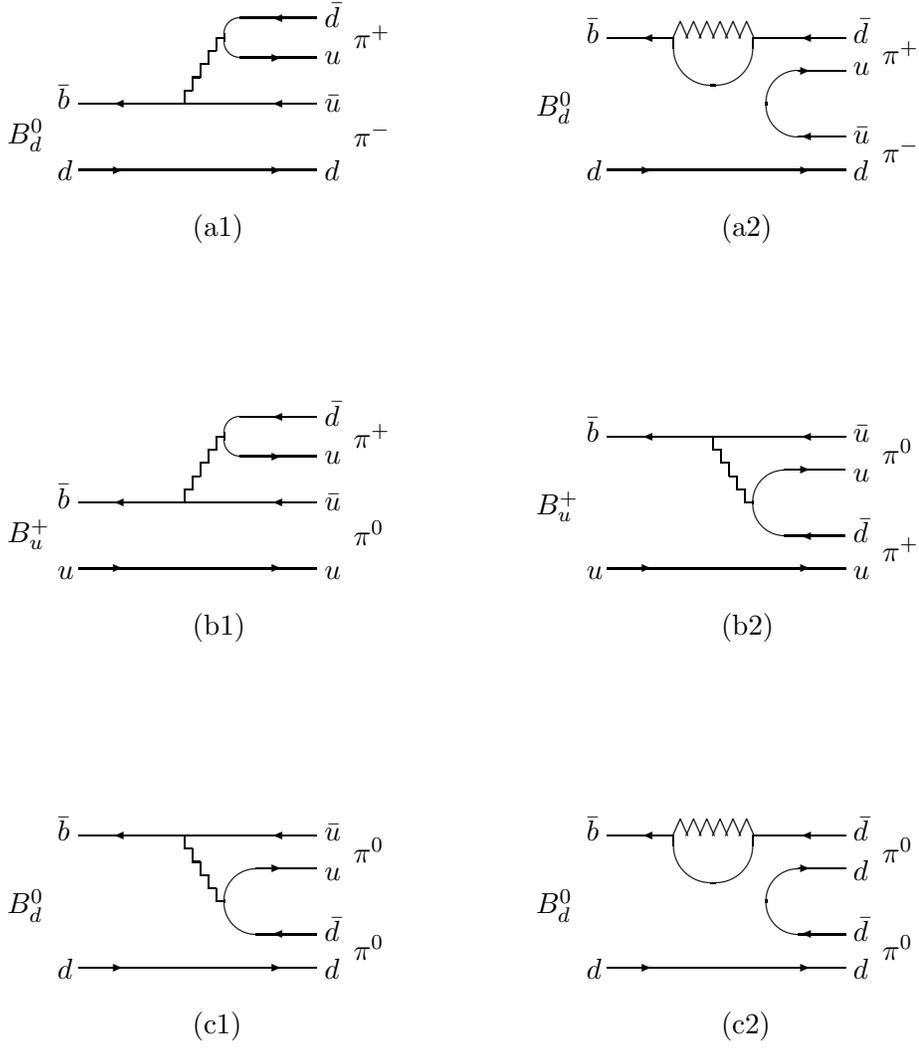

\end{document}